\documentclass[preprint,showpacs,preprintnumbers,amsmath,amssymb,nofootinbib]{revtex4}
\usepackage{graphicx}
\usepackage{dcolumn}
\usepackage{bm}
\usepackage{epstopdf}

\begin{document}
\title{Flutter of a Flag}
\author{M\'ed\'eric Argentina \footnote{Email: med@deas.harvard.edu}}
\author{L. Mahadevan \footnote{Email: lm@deas.harvard.edu}}
\affiliation{Division of Engineering and Applied Sciences, Harvard University,
Pierce Hall, 29 Oxford St., Cambridge, MA 02138 }
\date{20 February, 2004}
\begin{abstract}
We give an explanation for the onset of wind-induced flutter in a flag. Our theory accounts for the various physical mechanisms at work:  the finite length and the small but finite bending stiffness of the flag, the unsteadiness of the flow, the added mass effect and vortex shedding from the trailing edge. Our analysis allows us to predict a critical speed for the onset of flapping as well as the frequency of flapping. We find that in a particular limit corresponding to a low density fluid flowing over a soft high density flag, the flapping instability is akin to a resonance between the mode of oscillation of a rigid pivoted airfoil in a flow and a hinged-free elastic filament vibrating in its lowest mode.  

\end{abstract} 
\pacs{Valid PACS appear here}
\maketitle

The flutter of a flag in a gentle breeze, or the flapping of a sail in a rough wind are commonplace and familiar observations of  a rich class of problems involving the interaction of fluids and structures, of wide interest and importance in  science and engineering \cite{Paidoussis}.   Folklore attributes this flapping instability to some combination of (i)  the B\'{e}nard- von K\'arm\'an vortex street that is shed from the trailing edge of the flag, and (ii) the flapping instability to the now classical Kelvin-Helmholtz problem of  the growth of perturbations at an interface between two inviscid fluids of infinite extent moving with different velocities \cite{Rayleigh}. However a moment's reflection makes one realize that neither of these is strictly correct.  The frequency of vortex shedding from a thin flag (with an audible acoustic signature) is much higher than that of the observed flapping, while the initial differential velocity profile across the interface to generate the instability, the finite flexibility and length of the flag make it  qualitatively different from the Kelvin-Helmholtz problem. Following the advent of high speed flight, these questions were revisited in the context of aerodynamically induced wing flutter by Theodorsen \cite{Theodorsen}. While this important advance made it possible to predict the onset of flutter for rigid plates, these analyses are not directly applicable to the case of a spatially extended elastic system such as a  flapping flag.  Recently, experiments on an elastic filament flapping in a flowing soap film \cite{Zhang},  and of paper sheets flapping in a breeze \cite{Watanabe} have been used to further elucidate aspects of the phenomena such as the inherent bistability of the flapping and stationary states, and a characterization of the transition curve.  In addition, numerical solutions of the inviscid hydrodynamic (Euler) equations using an integral equation approach \cite{FitPope} and of the viscous (Navier-Stokes) equations \cite{Peskin} have shown that it is possible to simulate the flapping instability. However, the physical mechanisms underlying the instability remain elusive. In this paper, we aim to remedy this using the seminal ideas of Theodorsen \cite{Theodorsen}.  

\begin{figure}[ht]
	\includegraphics{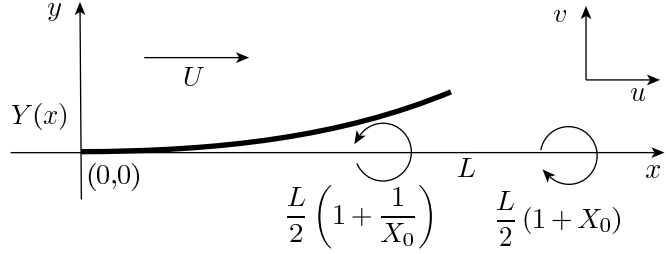}
	\caption{Schematic representation of the system. An elastic filament of length $L$, clamped at the origin is embedded in a 2-dimensional flow of an inviscid fluid with  velocity $U$ in the $x$ direction. Its lateral position is denoted by $Y(x,t)$. }
	\label{setup}
\end{figure}

We will start by considering the dynamics of an inextensible  one-dimensional elastic filament  of length $L$  and diameter $d$ and made of a material of density $\rho_s$ and Young's modulus $E$ embedded in a two dimensional   parallel flow of an ambient fluid with a density $\rho$ and kinematic viscosity $\nu$, shown schematically in Fig. \ref{setup} \footnote{Our analysis also carries over to the case of an elastic sheet a 3-dimensional parallel flow with no variations in the direction perpendicular to the main flow.}. We assume that the leading edge of the naturally straight filament is clamped at the origin with its tangent along the $x$ axis, and that far from the filament, the fluid velocity ${\bf U}=U {\bf x}$.  Then the transverse position of the filament $Y(x,t)$ satisfies the equation of motion \cite{LandauElasticity}:
\begin{equation}
	m Y_{tt}=-B Y_{xxxx}+ l \Delta P. \label{elasticity}
\end{equation}
Here, and elsewhere $A_b \equiv \partial A/\partial b$, $m=\rho_s \pi d^2/4$ is the mass per unit length of the filament, $B=\pi Ed^4/64$ its flexural rigidity, $l$ is the thickness of the fluid film \footnote{In the experiments with filaments in soap films \cite{Zhang}, $l\ne d$. For a sheet $l=1$, $m$ is a mass per unit area and  $B$ is now the bending stiffness per unit length.} and $\Delta P$ the pressure difference  across the filament due to fluid flow. In deriving (\ref{elasticity}) we have assumed that the slope of the filament is small so that we can neglect the effect of any geometrical nonlinearities; these become important in determining the detailed evolution of the instability  but are not relevant in understanding the onset of flutter. For the case when the leading edge of the flag is clamped and the trailing edge is free, the boundary conditions associated with (\ref{elasticity}) are \cite{LandauElasticity}:
\begin{eqnarray}
	Y(t,0)=0, & &Y_{x}(t,0)=0,  \nonumber \\
	Y_{xx}(t,L)=0, & &Y_{xxx}(t,L)=0. \label{bcs}
\end{eqnarray}
 To close the system (\ref{elasticity},\ref{bcs}) we must evaluate the fluid pressure $\Delta P$ by solving the equations of motion for the fluid in the presence of the moving filament. We will assume that the flow is incompressible, inviscid and irrotational.   The omission of viscous effects is justified if the shear stress $\rho\sqrt{\frac{\nu U^3}{L}}$ induced by the Blasius boundary layer  \cite{LandauFluid} is small compared to the fluid pressure $\rho U^2$  far away from the filament or equivalently if the characteristic Reynolds number    $Re= U L/\nu \gg 1$. In typical experiments, since $Re \sim 10^5$, this condition is easily met. Then we may describe the unsteady fluid flow as a superposition of  a non-circulatory flow and a circulatory flow associated with vortex shedding, following the pioneering work of Theodorsen \cite{Theodorsen}. This allows us to respect Kelvin's theorem preserving the total vorticity of the inviscid system (which is always zero) by considering a vortex sheet in the fluid  and an image sheet of opposite strength that is in the filament. Both flows may be described by a velocity potential $\phi$ which itself may be decomposed into a non-circulatory potential $\phi_{nc}$ and a circulatory potential $\phi_\gamma$ with $\phi = \phi_{nc} + \phi_{\gamma}$.  Then $\phi$ satisfies the Laplace equation  $\nabla^n2\phi=0$ characterizing the two-dimensional  fluid velocity field $(u,v)=(\phi_x, \phi_y)$. 

For small deflections of the filament, the transverse velocity of the fluid $v$ varies slowly along the filament. Then we may  use a classical result from airfoil theory \cite{Milne} for an airfoil moving with a velocity $v=Y_t+UY_x$  to deduce the non-circulatory velocity potential along the filament as \cite{LandauFluid}
\begin{equation}
	\phi_{nc}=\sqrt{x\left(L-x\right)}
	\left[
	Y_t+U Y_x
	\right], \label{nc}
\end{equation}
To determine the jump in pressure due to the non-circulatory flow we use the linearized Bernoulli relation so that
\begin{eqnarray}
P_{nc} &=&-2\rho(\partial_t \phi_{nc}+U\partial_{x}\phi_{nc}) \\ \nonumber
 &=&\frac{\rho U(2x-L)}{\sqrt{x(L-x)}}\left(Y_t+U Y_x\right)+\sqrt{x\left(L-x\right)}\rho Y_{tt}.
\end{eqnarray}
Here we note that  the fluid added-mass effect \footnote{When the filament moves, fluid must also be displaced and the sheet behaves as if it had more inertia \cite{LandauFluid}} is characterized by the term proportional to $Y_{tt}$, and we have neglected terms of order $O(Y_{xt})$ and higher associated with very slow changes in the slope of the filament.

Kelvin's theorem demands that vorticity is conserved in an inviscid flow of given topology. Thus,  the circulatory flow associated with vortex shedding from the trailing edge requires a vorticity distribution in the wake of the airfoil and a (bound) vorticity distribution in the airfoil to conserve the total vorticity. If a point vortex shed from the trailing edge of the filament with strength $-\Gamma$ has a  position $\frac{L}{2}\left(1+{X_0}\right)$, $X_0\geqslant 1$, we must add a point vortex of strength $\Gamma$ in the interior of  the sheet at $\frac{L}{2}\left(1+\frac{1}{X_0}\right)$. This leads to a circulatory velocity potential  along the filament \cite{Theodorsen}
\[
	\phi_\Gamma=-\frac{\Gamma}{2\pi}\arctan\left(\frac{\sqrt{x(L-x)}\sqrt{x_0^2-1}}{\frac{L}{2}(1+x_0)-x x_0}\right),
\]
where $x_0=\frac{X_0+1/X_0}{2}$ characterizes  the non-dimensional {\it center of vorticity} which is at $(1+x_0)/2$. Therefore for a distribution of vortices of strength $\gamma$ defined by   $\Gamma=\gamma\frac{L}{2} dx_0$, the circulatory velocity potential is
\begin{equation}
	\phi_{\gamma}=-\frac{1}{2\pi}\frac{L}{2}\int_1^\infty \arctan\left(\frac{\sqrt{x(L-x)}\sqrt{x_0^2-1}}{\frac{L}{2}(1+x_0)-x x_0}\right) \gamma dx_0,
	\label{Phigamma}
\end{equation}
To calculate the pressure difference  due to the circulatory flow, we assume that the shed vorticity moves with the flow velocity $U$ in the flow so that  $\partial_t \phi_\gamma=\frac{2}{L}U\partial_{x_0}\phi_\gamma$ \footnote{This implies a neglect of any acceleration phase of the vorticity, a reasonable assumption at high $Re$.}. Then,  we may write \cite{Theodorsen}:
 \begin{equation}
 	P_\gamma=
	- \frac{\rho U}{2\pi \sqrt{x(L-x)}}
	\int_1^\infty\frac{2 x+L(x_0-1)}{\sqrt{x_0^2-1}}\gamma dx_0
	\label{PressureBeforeKutta}
 \end{equation}
The vortex sheet strength $\gamma$ in the previous expression is determined using the Kutta condition  which enforces the physically reasonable condition that  the horizontal component of the velocity does not diverge at the trailing edge \footnote{This is tantamount to the statement that that the inclusion of viscosity, no matter how small, will regularize the flow in the vicinity of the trailing edge.}:
\begin{equation}
	\partial_x\left(\phi_{\gamma}+\phi_{nc}\right)|_{x=L}={\mathrm{finite}}
	\label{KuttaCondition}
\end{equation}
Substituting (\ref{nc}, \ref{Phigamma}) into (\ref{KuttaCondition}) yields the relation
\begin{equation}
	\frac{1}{2\pi}\int_1^\infty \sqrt\frac{x_0+1}{x_0-1}\gamma dx_0=Y_t+U Y_{x}
	\label{KuttaGamma}
\end{equation}
 Multiplying and dividing (\ref{PressureBeforeKutta}) by the two sides of (\ref{KuttaGamma}) we obtain
\begin{eqnarray}
	P_\gamma=&
	- \frac{\left(L(2C-1)+2x(1-C)\right)}{ \sqrt{x(L-x)}}\rho U\left(Y_t+U Y_{x}\right)
\end{eqnarray}
where
\begin{eqnarray}
C[\gamma]=&\int_1^\infty\frac{x_0}{\sqrt{x_0^2-1}}\gamma dx_0/\int_1^\infty \sqrt\frac{x_0+1}{x_0-1}\gamma dx_0
	\label{Wagner}
\end{eqnarray}
is  the Theodorsen functional  \cite{Theodorsen} which  quantifies the unsteadiness of the flow.  For example, for an airfoil at rest which starts to move suddenly at velocity $U$,   $\gamma=\delta(x_0-\frac{2}{L}Ut)$ corresponding to the generation of lift due to a vortex that is shed and advected with the fluid. Then $C = (1+\frac{L}{2 t U})^{-1}$ and we see that as $Ut/L \rightarrow \infty, C \rightarrow 1$, which limit corresponds to the realization of  the Kutta condition for steady flow \cite{LandauFluid}. Adding up the contributions to the pressure jump across the filament from the circulatory and non-circulatory flows, we have  $\Delta P = P_{nc} + P_{\gamma}$, i.e.
\begin{equation}
	\Delta P=-\rho U C[\gamma] f\left(\frac{x}{L}\right)(Y_t+UY_{x})-L\rho n\left(\frac{x}{L}\right)Y_{tt}.
	\label{DeltaP}
\end{equation}
where the dimensionless functions $n(s)$ and $f(s)$  are 
\begin{eqnarray}
	f(s)&=&2\sqrt\frac{1-s}{s},\\
	n(s)&=&2\sqrt{(1-s)s}.
\end{eqnarray}
Substituting (\ref{DeltaP}) in (\ref{elasticity}) gives us a single equation of motion for the hydrodynamically driven filament
\begin{equation}
	\begin{array}{ll}
		m Y_{tt}=&-B Y_{xxxx}\\
		&-\rho U C[\gamma] f(\frac{x}{L})(Y_t+UY_{x})\\
		&-L\rho n\left(\frac{x}{L}\right)Y_{tt}.
	\end{array}
	\label{finalDimensioned}
\end{equation}
with $C[\gamma]$ determined by (\ref{Wagner}). We note that (\ref{finalDimensioned}) accounts for the unsteady flow past a filament of finite length unlike previous studies \cite{FitPope}, and thus includes the effects of vortex shedding and fluid added-mass. To make (\ref{finalDimensioned}) dimensionless, we scale all lengths with the length $L$ of the flag, so that $x=s L, Y=\eta L$, and scale time with the bending time $L/U_B$, where $U_B=\frac{1}{L}\sqrt\frac{B}{m}$ is the velocity of bending waves of wavelength $2 \pi L$. Then (\ref{finalDimensioned}) may be written as
\begin{equation}
		{\cal M}\eta_{\tau\tau}=-\eta_{ssss}
		-\mu \delta C[\gamma] f(s)\left(  \eta_\tau+\delta \eta_{s}\right)
		\label{final}
\end{equation}
Here ${\cal M}=1+ \mu n(s)$ where $\mu=\frac{l\rho L}{m}=\frac{4\rho}{\pi \rho_s} \frac{L l}{d^2}$  characterizes the added mass effect  and the parameter $\delta =\frac{U}{U_b}$ is the ratio of the fluid velocity to the bending wave velocity in the filament.  We can use symmetry arguments to justify the aerodynamic pressure $C[\gamma] f(s)\left(  \eta_\tau+\delta \eta_{s}\right)$: the term $\eta_s$ arises because the moving fluid breaks the $s\rightarrow-s$ symmetry, while the term $\eta_\tau$ arises because  the filament exchanges momentum with the fluid, so that the time reversibility $\tau \rightarrow-\tau$ symmetry is also broken.  These two leading terms  in the pressure, which could have  been written down on grounds of symmetry, correspond  to a lift force  proportional to $\eta_s$, and a  frictional damping proportional to $\eta_\tau$. By considering the detailed physical mechanisms, we find that the actual form of these terms is more complicated due to the inhomogeneous dimensionless functions $f(s), n(s)$. Thus, understanding the flapping instability reduces to a stability analysis of the trivial solution $\eta=0$ of the system (\ref{final},\ref{bcs}) and the determination of a transition curve as a function of the problem parameters $\mu,\delta$.

\begin{figure}[ht]
	\includegraphics{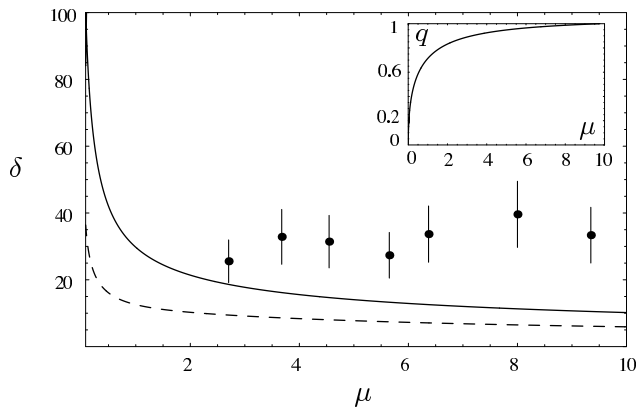}
	\caption{Stability diagram for (\ref{final},\ref{bcs}) as a function of the added mass parameter $\mu$ and the scaled flow velocity $\delta$. The solid line represents the transition curve when vortex shedding is taken into account, i.e. $C \neq 1$. The dashed line represents the transition curve using the quasi-steady approximation where $C=1$. In the inset we show the dimensionless wavenumber of the instability $q=\frac{\omega}{2\delta}$  as a function of $\mu$. When $\mu\ll 1$, $q$ tends to be zero and $C(q)\rightarrow 1$. The dots correspond to experimental data characterizing the transition to flutter in three-dimensional flows past flexible sheets of paper \cite{Watanabe}; the large error bars are a consequence of the variations due to three-dimensional effects as well as regions of bistability where both the flapping and stationary state are stable.
	}
	\label{QuasiSteady}
\end{figure}

\begin{figure}[thb]
	\includegraphics[width=\columnwidth]{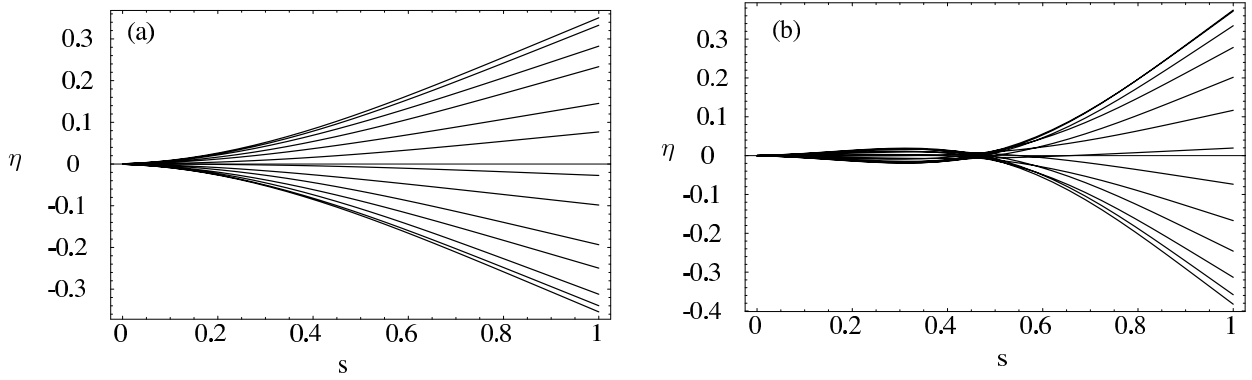}
	\caption{Snapshots of the deflection of the filament $\eta$ at the instability threshold for (a) $\mu=0.2, \delta \approx 66$ and  (b) $\mu=25, \delta \approx 6.6$.}
	\label{EigenVectors}
\end{figure}

Since the free vortex sheet is advected with the flow, the vorticity distribution may be written as $\gamma=\gamma(\frac{2 U}{L}(t-t_1)-x_0)$, with $(1+x_0)/2$ denoting the {\it center of vorticity}, $t_1$ being the time at which shedding occurs; in dimensionless terms reads $\gamma=\gamma(2\delta(\tau-\tau_1)-x_0)$. Accounting for the oscillatory nature of the flapping instability with an unknown frequency $\omega$ suggests that an equivalent description of  the vorticity distribution is given by $\gamma= A e^{i(\omega (\tau-\tau_1)-q x_o)}$ where $q=\omega/2\delta$ is a non dimensional wave number of the vortex sheet.  Using the above traveling wave form of the vorticity distribution  in (\ref{Wagner}) we get an expression for the Theodorsen function \cite{Theodorsen} 
\begin{equation}
	C[\gamma]=C(q)=\frac{H_1(q)}{H_0(q)+i H_1(q)}, 
	\label{Cn}
\end{equation}
where $H_i$ are Hankel functions of $ith$ order.  Substituting the separable form $\eta(s,\tau) =\xi(s) e^{\sigma \tau}$   into (\ref{final}) we get:

\begin{equation}
 	\sigma^2{\cal M}\xi=
	-\xi_{ssss}
	-C[\gamma]\mu\delta f(s) \left(\sigma \xi+\delta \xi_s \right) 
	\label{Cne1}
\end{equation}
At the onset of the oscillatory instability, $Re(\sigma)=0$, so that $\sigma=i\omega$ and $C[\gamma]$ is given by (\ref{Cn}). Then (\ref{Cne1}, \ref{bcs}) constitutes a nonlinear eigenvalue problem for  $\omega, \xi$ given the nonlinear $\omega$ dependence of the Theodorsen function $C(q)=C(\omega/2\delta)$  in (\ref{Cn}). We solve the resulting system numerically with the AUTO package \cite{auto}, using a continuation scheme in $\omega$ starting with a guess for the Theodorsen function $C(\omega/2\delta)=C(0)=1$ . As we shall see later,  this limit corresponds to the quasi-steady approximation  \cite{Fung}. In Fig. \ref{QuasiSteady} we show the calculated transition curve;  when $\delta>\delta_c(\mu)$, ${\mathrm {Re}}(\sigma)>0$ with ${\mathrm{Im}}(\sigma)\neq 0 $, i.e. an oscillatory instability leading to flutter arises. We see that for sufficiently large $\delta$ the filament is always unstable,  i.e. large enough fluid velocities will always destabilize the elastic filament. As $\mu \gg 1$, the added mass effect becomes relatively more important and it is easier for the higher modes of the filament to be excited. In Fig. \ref{EigenVectors} we show the mode shapes when $\mu <1$ and $\mu \gg 1$; as expected the most unstable mode for $\mu \gg 1$ is not the fundamental mode of the filament. We also see that the normalized amplitude of the unstable modes  is maximal at the trailing edge; this is a consequence of the inhomogeneous functions $f(s), n(s)$ in (\ref{final})  as well as the clamped leading edge and a free trailing edge.

To further understand the instability, we now turn to a simpler case using the quasi-steady approximation \cite{Fung}. This supposes that the lift forces are slaved adiabatically to those on a stationary airfoil  with the given instantaneous velocity $Y_t + UY_x$, so that $C=1$. By assuming that the Kutta condition is satisfied instantaneously, we over-estimate the lift forces and thus expect to get a threshold for stability that is slightly lower than if $C \ne 1$. To characterize the instability in this situation, we substitute  an inhomogeneous perturbation of the form $\eta(s,\tau) =\xi(s) e^{\sigma \tau}$ into  (\ref{final},\ref{bcs}) and solve the resulting eigenvalue problem to determine the growth rate $\sigma=\sigma(\delta,\mu)$. In Fig. 2, we show the stability boundary corresponding to the quasi-steady
approximation.    We note that the stability boundary when $C\neq1$ accounting for vortex shedding corresponds to a higher value of the scaled fluid velocity $\delta$ than that obtained using the quasi-steady approximation $C=1$, and is a consequence of the quasi-steady approximation which over-estimates the lift forces. 

When $\mu\ll 1$, corresponding to either a  fluid of very low density or a filament of very high density, Fig. \ref{QuasiSteady} shows that the corresponding instability occurs for high fluid velocities $ U\gg U_B$. Then $q \rightarrow 0$, as confirmed in the inset to Fig. \ref{QuasiSteady}. Therefore $C(q) =C(0) =1$ so that in this limit the quasi-steady hypothesis is a good approximation. In the limit $\mu\rightarrow0$, we must have $\mu\delta^2=const$ so that the aerodynamic pressure which drives the instability remains finite. Then the system ({\ref{final}) becomes Hamiltonian \footnote{This is because the term breaking time reversal symmetry $\mu \delta \eta_\tau$ becomes negligibly small.} and may be written as:
  \begin{figure}[tb]
	\includegraphics[width=0.65\columnwidth]{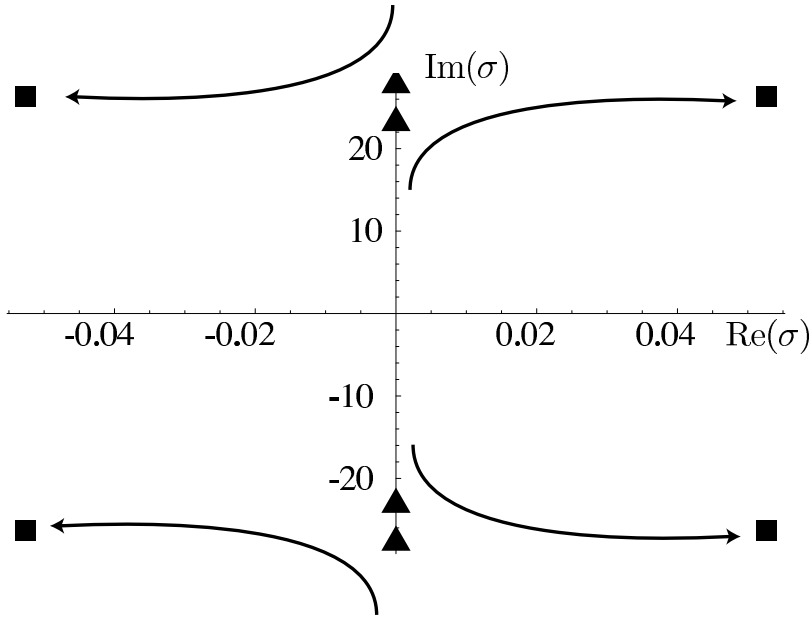}
	\caption{Spectrum $Im(\sigma), Re(\sigma)$ of the Hamiltonian system (\ref{hamiltonian},\ref{bcs}) when $\mu\ll1$ (with $\delta_c=\frac{10.08}{\sqrt{\mu}}$). The eigenvalues with the smallest absolute value are  plotted  for $\delta=0.9\delta_c$ (triangle) and  and for  $\delta=1.1\delta_c$ (square). We see that instability occurs via a collision and splitting of two pairs of eigenvalues along the imaginary axis (indicated by the arrows), and is a signature of a 1:1 resonance or a Hamiltonian Hopf bifurcation.}
	\label{Resonance}
\end{figure}

\begin{equation}
		\eta_{\tau\tau}=-\eta_{ssss}
		-\mu \delta^2 f(s)\eta_{s}
		\label{hamiltonian}
\end{equation}
The two terms on the right hand side  of (\ref{hamiltonian}) correspond to the existence of two different modes of oscillation: (i) that of a flexible filament bending with a frequency that is dependent on the wavenumber and (ii) that of a rigid filament in the presence of flow-aligning aerodynamic forces. In this limiting case, we can clearly see the physical mechanisms at work in determining the stability or instability of the filament: small filaments are very stiff in bending, but as the filament length becomes large enough for the fluid pressure to excite a resonant bending instability the filament starts to flutter. Equivalently, the instability is observed when the bending oscillation frequency become of the order of the frequency of oscillations of a hinged rigid plate immersed in a flow. To see this quantitatively, we look for solutions to  (\ref{hamiltonian},\ref{bcs})  of the form $\eta(s,\tau) = \xi(s) e^{\sigma \tau}$ and compute the associated spectrum $\sigma(\delta)$. In Fig. \ref{Resonance}, we show that for $\delta<\delta_c=10.08/\sqrt{\mu} $ with $\mu \ll 1$, the spectrum lies on the imaginary axis as expected, and as $\delta \ge \delta_c$, the four eigenvalues with smallest absolute value collide and split,  leading to an instability via a Hamiltonian Hopf Bifurcation  or a 1:1 resonance \cite{Marsden}. 

As $\mu \sim O(1)$, the effective damping term $\mu \delta C f(s)\eta_{\tau}$ becomes important, so that the spectrum is shifted to the left, i.e. $Re(\sigma)<0$. In this case, although the instability is not directly related to a resonance, the physical mechanism remains the same, i.e. a competition between the destabilizing influence of part of the fluid inertia and the stabilizing influence of elastic bending, subject to an effective damping due to fluid motion. This simple picture allows us to estimate the criterion for instability  by balancing the bending forces $\frac{B \xi}{L^4}$ with the aerodynamic forces $l\rho U^2 \frac{\xi}{L}$ so that for a given flow field the critical length of the filament above which it will flutter  is
\begin{equation}
 	L_c\sim\left(\frac{B}{l \rho U^2}\right)^{1/3}, \label{L_c}
\end{equation}
which in dimensionless terms corresponds to $\delta \sim 1/\mu^{1/2}$. Then the typical flapping frequency $\omega$ is given by balancing filament inertia $m \omega^2\frac{\xi}{L}$ with the aerodynamic forces $l\rho U^2 \frac{\xi}{L}$ and leads to
\begin{equation}
 	\omega \sim\sqrt{\frac{l \rho U^2}{m L}}. \label{omega}
\end{equation}
Using  typical experimental parameters values from experiments \cite{Zhang}, we find that $L_c \sim 0.2\ cm$ with a frequency $\omega/2 \pi=89\  Hz$ in qualitative agreement with the experimentally observed values $L_c=4\ cm$ and $\omega/2\pi \sim 50\  Hz$. In Fig. \ref{QuasiSteady}, we also show the experimental transition curve obtained from a recent study on the onset of flutter in paper sheets \cite{Watanabe}. The large error bars in the experimental data are due to the fact that there is  a region of bistability wherein both the straight and the flapping sheet are stable. Our linearized theory cannot capture this bistability without accounting for the various possible nonlinearities in the system arising from geometry. But even without accounting for these nonlinearities, there is a systematic discrepancy between our theory  and the data which consistently show a higher value of $\delta$ for the onset of the instability.  While there are a number of possible reasons for this, we believe that there are two likely candidates: the role of three-dimensional effects and the effect of the tension in the filament induced by the Blasius boundary layer, both of which would tend to stabilize the sheet and thus push the onset to higher values of $\delta$.   

Nevertheless our hierarchy of models starting with the relatively simple Hamiltonian picture to the more sophisticated quasi-steady and unsteady ones have allowed us to dissect the physical mechanisms associated with flapping in a filament with a finite length and finite bending stiffness and account for the added-mass effect,  the unsteady lift forces and vortex shedding.  They also provide a relatively simple criteria for the onset of the instability in terms of the scaling laws (\ref{L_c}, \ref{omega}).    Work currently in progress includes a detailed comparison with a two-dimensional numerical simulation  and  will be reported elsewhere \cite{AM2004}. 
 
\small {Acknowledgments: Supported by the European Community through the Marie-Curie Fellowship HPMF-2002-01915 (MA) and the US Office of Naval Research through a Young Investigator Award (MA, LM).}

\end{document}